\documentclass[conference]{IEEEtran}
\usepackage{cite}
\def\BibTeX{{\rm B\kern-.05em{\sc i\kern-.025em b}\kern-.08em
    T\kern-.1667em\lower.7ex\hbox{E}\kern-.125emX}}

\usepackage[numbers,sort&compress]{natbib}

\usepackage[dvipsnames]{xcolor}

\ifCLASSINFOpdf
   \usepackage[pdftex]{graphicx}
   \usepackage{subfigure}
   \graphicspath{{Figures/}{./}}
\else
\fi
\usepackage{amsmath}
\usepackage{amssymb}

\usepackage{array}
\begin{document}
%
\title{Mobility Performance Analysis of \\ Scalable Cell-Free Massive MIMO}
%
%
%

        
\author{\IEEEauthorblockN{Yunlu~Xiao,
Petri~Mähönen and
Ljiljana~Simić}
\IEEEauthorblockA{Institute for Networked Systems, 
RWTH Aachen University,
Aachen, Germany\\ Email: \{yxi, pma, lsi\}@inets.rwth-aachen.de}}

\maketitle

\begin{abstract}
While scalable cell-free massive MIMO (\mbox{CF-mMIMO}) shows advantages in static conditions, the impact of its changing serving access point (AP) set in a mobile network is not yet addressed. In this paper we first derive the CPU cluster and AP handover rates of scalable CF-mMIMO as exact numerical results and tight closed form approximations. We then use our closed form handover rate result to analyse the mobility-aware throughput. We compare the \mbox{mobility-aware} spectral efficiency (SE) of scalable CF-mMIMO against distributed MIMO with pure network- and UE-centric AP selection, for different AP densities and handover delays. Our results reveal an important trade-off for future dense networks with low control delay: under moderate to high mobility, scalable CF-mMIMO maintains its advantage for the $\mathbf{{95}^{th}}$-percentile users but at the cost of degraded median SE.

\end{abstract}


%
\IEEEpeerreviewmaketitle

\section{Introduction}
%
%
%
%

Cell-free massive multiple-input multiple-output (\mbox{CF-mMIMO}) is proposed in \cite{cfvs} as a means of providing superior performance to the traditional cellular system, by combining the benefits of massive and distributed MIMO. Namely, in \mbox{CF-mMIMO} \textit{all} access points (APs) in the network jointly serve a given user equipment (UE), thus resolving the traditional cell-edge effect for significantly more uniform performance over the network. Since \mbox{CF-mMIMO} thus exhibits no cell boundaries, there is no change in the serving AP set when the UE moves, and handovers are also no longer an issue. However, in practice, managing all APs under one single central processing unit (CPU) and coherently processing data for all APs to simultaneously serve all UEs is not scalable for realistically large networks, since the complexity and resource requirements in signal processing and signalling tend towards infinity for a large number of UEs \cite{scalableCF}. Therefore, a scalable \mbox{CF-mMIMO} network is proposed in \cite{cfsim}, where a given UE is still served by a limited subset of APs. Specifically, the UE first selects a few APs that have the best channel conditions, and is then served by the disjoint CPU clusters that those APs belong to. Scalable \mbox{CF-mMIMO} is thus a form of distributed MIMO, where the finite serving AP set changes as the UE moves through the network, reintroducing handovers and related mobility overheads.

The AP serving set in scalable \mbox{CF-mMIMO} is effectively formed by the APs belonging to CPU clusters that surround the UE. The AP selection method proposed in \cite{cfsim} is thus a mix of \textit{network-centric} and \textit{UE-centric} approaches, and we refer to it as the \textit{hybrid} method for distributed MIMO AP selection. Compared to conventional coordinated multipoint with joint transmission (CoMP-JT) corresponding to \mbox{network-centric} distributed MIMO, where the UE is only served by the CPU cluster it resides in \cite{comp}, the hybrid method provides better performance for cluster-edge UEs. On the other hand, since APs are still managed by disjoint CPU clusters, the hybrid method is assumed to have lower control overhead than the pure UE-centric (PUE) cooperation framework \cite{pue}. However, while the hybrid method shows these advantages in a static network, the mobility aspect has not yet been addressed. When UE mobility is taken into account, the serving AP set in a distributed MIMO network will change when the UE moves, incurring mobility costs such as handover delays and signalling overheads \cite{optimizing}. Moreover, the link unavailability during handover will impact the network throughput, resulting in low mobility throughput performance for high handover rate AP selection methods \cite{mobilityaware}. Whereas the performance of PUE and CoMP-JT under mobility has been studied \cite{tutorial, optimizing}, a mobility analysis of scalable CF-mMIMO with its hybrid AP selection method is missing from the literature. Importantly, it is thus not yet clear whether scalable \mbox{CF-mMIMO} in fact outperforms distributed MIMO with other AP selection methods, i.e. CoMP-JT and PUE, once mobility is taken into account.

To address this gap, we present two main contributions. First, we use stochastic geometry analysis to derive the CPU cluster and AP handover rates for the hybrid method as exact numerical expressions and tight closed form approximations, and confirm their accuracy via simulations. Second, we use our closed form handover rate result to analyse the mobility performance in terms of \mbox{mobility-aware} spectral efficiency (SE). We compare the mobility-aware SE of scalable \mbox{CF-mMIMO} against distributed MIMO with \mbox{CoMP-JT} and PUE AP selection, for different AP densities and handover delays. Our results reveal an important trade-off for future dense networks with low control delay: scalable \mbox{CF-mMIMO} with hybrid AP selection outperforms CoMP-JT for \mbox{``cluster-edge''} UEs, but at the cost of reduced median SE, even at moderate mobility.

This paper is organized as follows. Sec. \ref{systemmodel} presents our system model. Sec. \ref{H} presents our handover rate analysis for the hybrid AP selection method of scalable \mbox{CF-mMIMO}. Sec. \ref{SEmo} details our mobility-aware UE throughput model. Sec. \ref{sim} presents our results and Sec. \ref{conclude} concludes the paper.

\begin{figure*}[!t] 
	\centering  
	\subfigure[CoMP-JT]{
		\label{Nposition.sub.1}
		\includegraphics[width=0.16\linewidth]{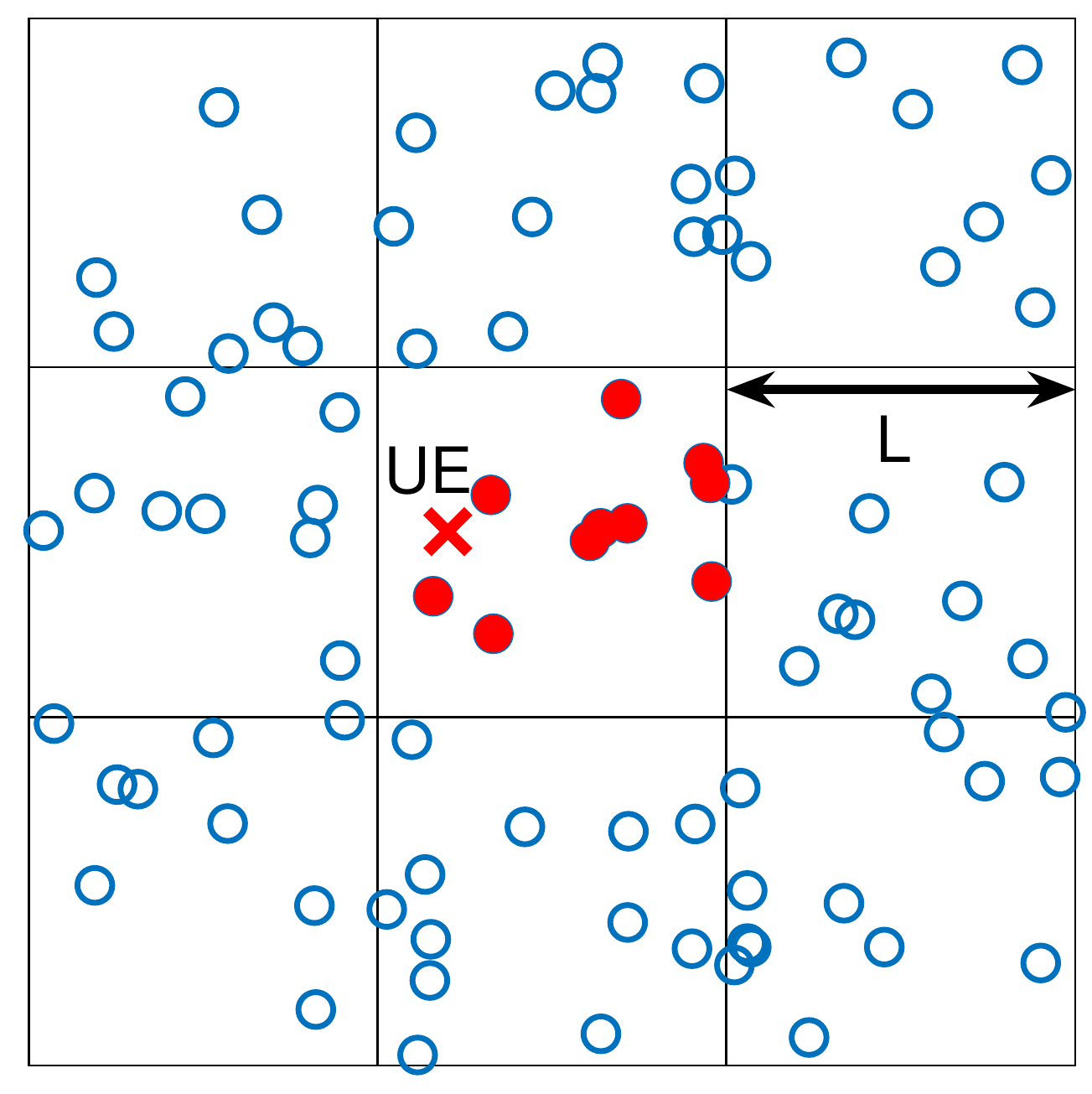}}	
	\subfigure[PUE, $K=20$]{
		\label{Nposition.sub.2}
		\includegraphics[width=0.16\linewidth]{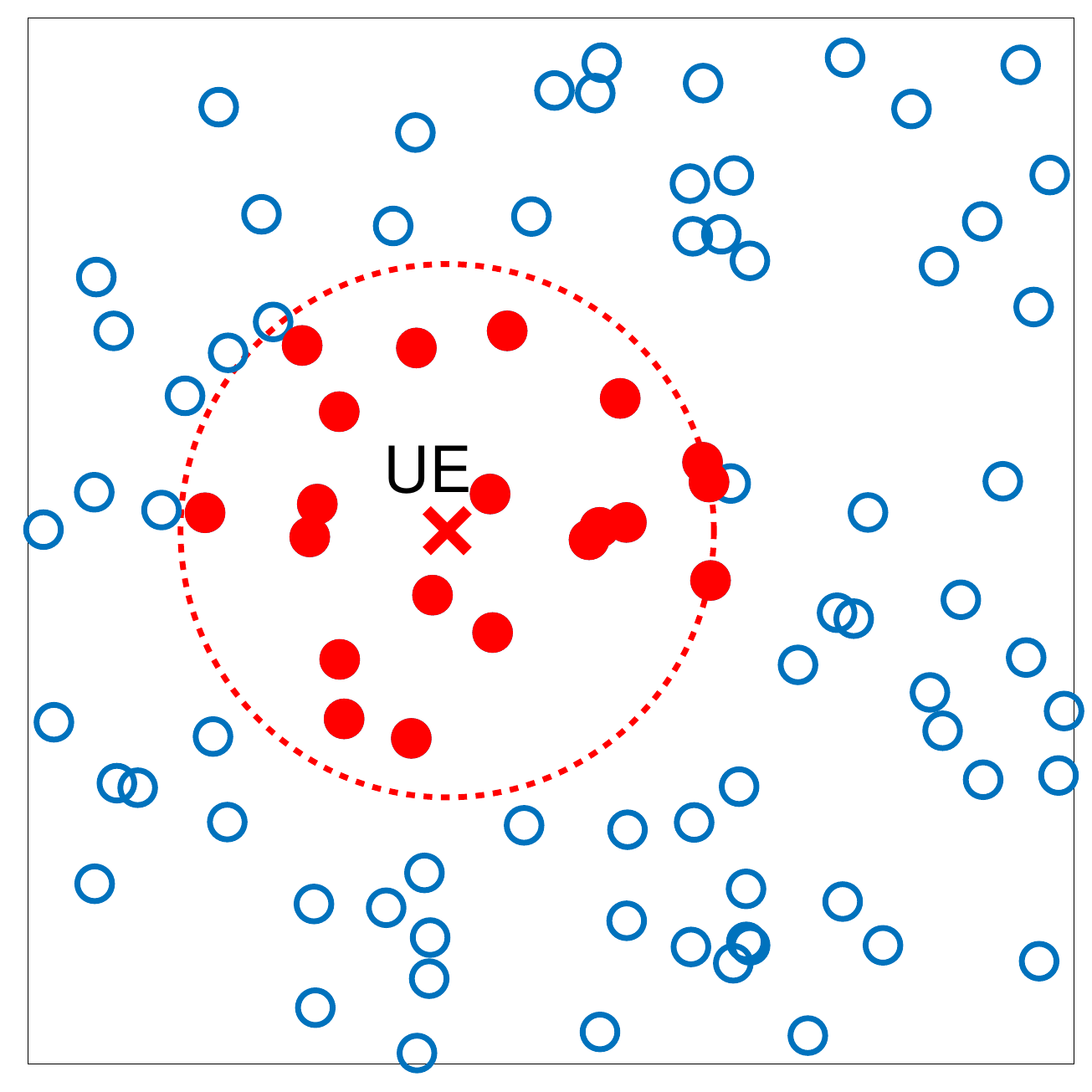}}
	\subfigure[hybrid, $K=5$]{
		\label{Nposition.sub.3}
		\includegraphics[width=0.16\linewidth]{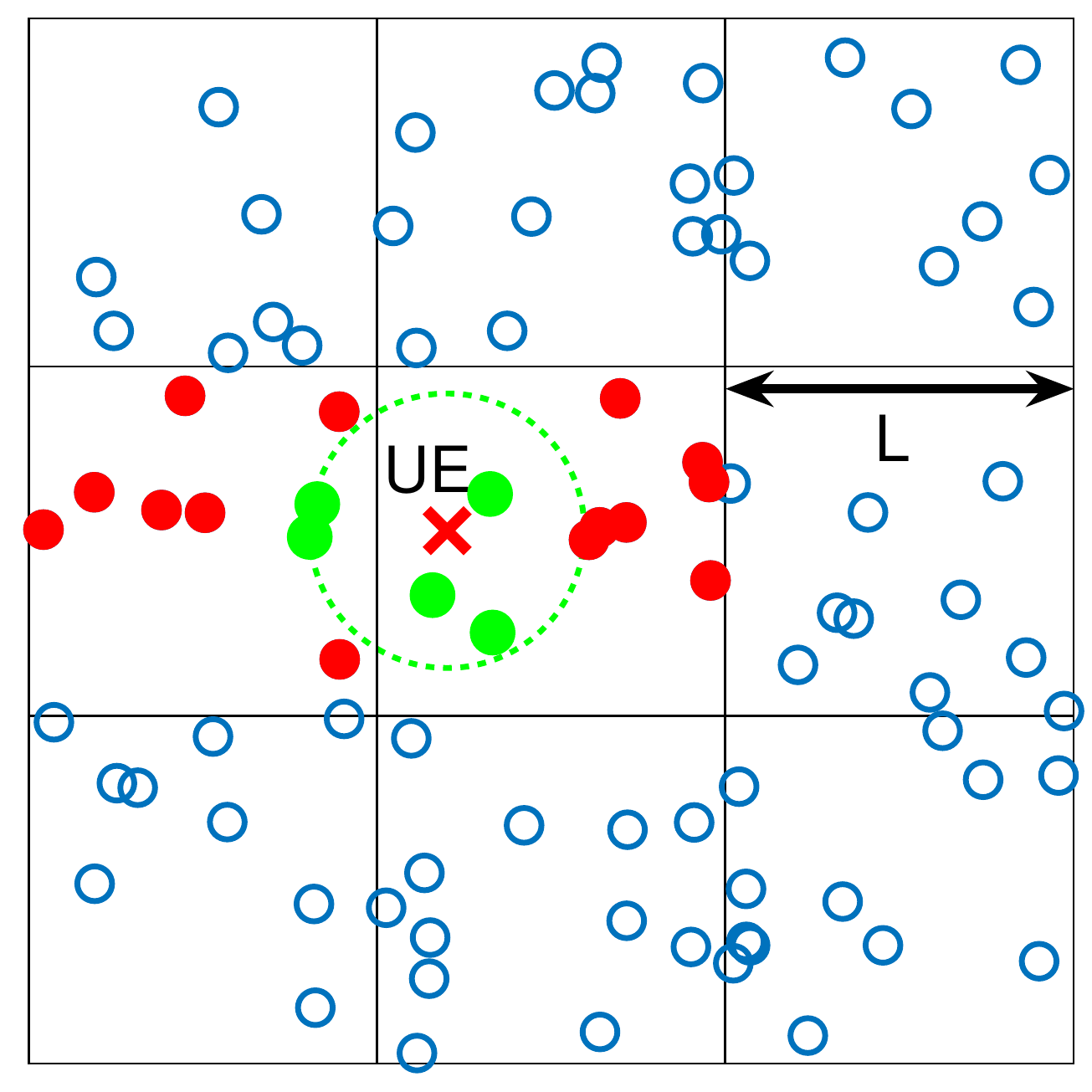}}
	\caption{Serving set for different AP selection methods for an example UE (red cross). Black lines represent the border of CPU cluster areas of length $L$ and $Q=10$, dashed circles represent the closest AP group, small solid circles represent APs, and filled circles represent the selected serving APs.}
	\label{Nposition}
\end{figure*}
\section{System Model}
\label{systemmodel}

\subsection{Distributed MIMO Network Model}
We consider the distributed MIMO system model in \cite{cfvs} and its scalable version in \cite{cfsim}. All APs and UEs in the network are equipped with a single antenna and scattered with density $\lambda$ and $\lambda_U$, respectively, in \mbox{two-dimensional} Euclidean space according to a homogeneous Poisson point process. We focus on the system's downlink performance and the impact of handover delay in this paper. The closed form expression in Eq. (24) in \cite{cfvs} is used to analyse the per-UE downlink throughput performance, while the power control scheme for CF-mMIMO is detailed in \cite{cfsim}. The underlying distributed MIMO model also assumes a \mbox{three-slope} log-distance path loss model, conjugate beam-forming, MMSE channel estimation, and non-orthogonal uplink pilots \cite{cfvs, cfsim}.

\subsection{AP Selection Method}
In addition to scalable CF-mMIMO with its hybrid method in \cite{cfsim}, we consider for comparison two more AP selection methods to form the serving set of a UE in the distributed MIMO network: conventional CoMP-JT in \cite{comp} and the PUE method in \cite{pue}. The handover rate of the hybrid method will be derived in Sec. \ref{H} and the mobility performance of the methods will be compared in Sec. \ref{SEmo}. The serving set for each AP selection method, as illustrated in Fig. \ref{Nposition}, is formed as follows. 

For CoMP-JT, APs are assigned to disjoint CPU clusters based on their geographic locations. On average there are $Q$ APs in one CPU cluster. The APs in one CPU cluster connect to the CPU via fronthaul links, while all CPUs connect to the core network via backhaul links. The UE is then served by the APs in its CPU cluster area, which we model as a square with length $L$ (\textit{cf}. Fig. \ref{Nposition.sub.1}, where $Q=10$). 

For PUE, all APs are managed by the common control plane via backhaul links instead of clusters. The UE is served by the $K$ APs with the best channel condition; in this paper we choose the $K$ closest APs to the UE as in \cite{optimizing} since for the assumed log-distance path loss model this is equivalent to the $K$ APs with the strongest received power level. Fig. \ref{Nposition.sub.2} shows an example with $K=20$. 

For the hybrid method, APs are also assigned to disjoint CPU clusters as in CoMP-JT. The UE first selects the $K$ closest APs, and is then served by all APs in the CPU clusters that the $K$ closest APs belong to. In the example in Fig. \ref{Nposition.sub.3}, the UE is served by all APs in two CPU clusters of the $K=5$ closest APs shown in green.

\subsection{Mobility Model}
We consider static APs and observe the movement of a reference UE assuming the random way point mobility model \cite{towards}. In order to derive the closed form expression for handover rate, we assume that in each moving period, the UE moves with a fixed speed $v$ and a random direction $\theta$, where the transition lengths of all moving period are Rayleigh distributed. The mobility-aware throughput model is given in Sec. \ref{SEmo}.

\section{Handover Rate Analysis for the \\ Hybrid AP Selection Method}
\label{H}
In this section, we derive the exact numerical expression and its closed form approximation for both the cluster and AP handover rates of the hybrid AP selection method of scalable CF-mMIMO, using the stochastic geometric and virtual cell analytical framework proposed in \cite{optimizing} and \cite{disjoint}. To derive the handover rate for a UE that moves with speed $v$ in a time period $\delta t$ in area $A$, we apply the Buffon’s needle problem to calculate the probability that this UE crosses the cell boundaries. That is, we find the probability that a needle with length $v \delta t$ crosses a line, where the lines are the cell boundaries \cite{tutorial}. In order to obtain this probability, we first characterize the handover boundary set of the hybrid method and derive the length intensity of this boundary set,  which is in turn derived through the area intensity.

\subsection{Handover Boundary}
We characterize the handover boundary of the hybrid method as a modified $K^{th}$-order Poisson Voronoi cell boundary set, defined as
\begin{equation}
\begin{aligned}
\mathbf{T}_{0K}^{(1)}=&\left\{\mathbf{y} \in \mathbb{R}^{2} \mid \exists\left\{\mathbf{x}_{1}, \mathbf{x}_{2}, \ldots, \mathbf{x}_{K-1}, \mathbf{x}_{K}, \mathbf{x}_{K}^{\prime}\right\} \subset \Phi\right.\\
&\text { s.t. }|\mathbf{z}-\mathbf{y}| \leq \left|\mathbf{x}_{K}-\mathbf{y}\right|=\left|\mathbf{x}_{K}^{\prime}-\mathbf{y}\right| \leq|\mathbf{x}-\mathbf{y}|, \\
& \mathbf{SQ({x}_{K})\neq SQ({x}_{K}^{\prime})}, 
\forall \mathbf{z} \in \left\{\mathbf{x}_{1}, \ldots, \mathbf{x}_{K-1}\right\} \text { and } \\
& \forall \mathbf{x} \in \left.\Phi \backslash\left\{\mathbf{x}_{1}, \ldots, \mathbf{x}_{K-1}, \mathbf{x}_{K}, \mathbf{x}_{K}^{\prime}\right\}\right\},
\end{aligned}
\label{T1}
\end{equation}

\noindent where $\Phi$ is the set of all APs, $ \mathbf{x}_{K}$ and $\mathbf{x}_{K}^{\prime}$ are the two $K^{th}$ closest AP locations and $\mathbf{SQ({x}_{K})}$ is the CPU cluster that point $\mathbf{x_K}$ belongs to. If a point $\mathbf{y}$ is in $\mathbf{T}_{0K}^{(1)}$, then the distances between $\mathbf{y}$ and the two $K^{th}$ closest AP are the same. This distance is greater than or equal to the distances to the other closest APs $(|\mathbf{z}-\mathbf{y}|)$, but is less than or equal to the distances to all the APs outside the closest set $(|\mathbf{x}-\mathbf{y}|)$. Also, the two APs should be in different CPU clusters which makes $\mathbf{T}_{0K}^{(1)}$ a subset of the $K^{th}$-order Poisson Voronoi cell boundary. When the UE crosses $\mathbf{T}_{0K}^{(1)}$, both the closest AP set and the CPU cluster that those APs belong to will change, leading to a handover of one CPU cluster. The number of CPU cluster handovers will be the number of intersections between the UE trajectory and $\mathbf{T}_{0K}^{(1)}$. The CPU cluster handover rate will be the number of handovers in one unit time, and the AP handover rate will be that scaled by the number of APs per CPU cluster.

\begin{figure}[!t] 
	\centering
	\includegraphics[width=0.9\columnwidth]{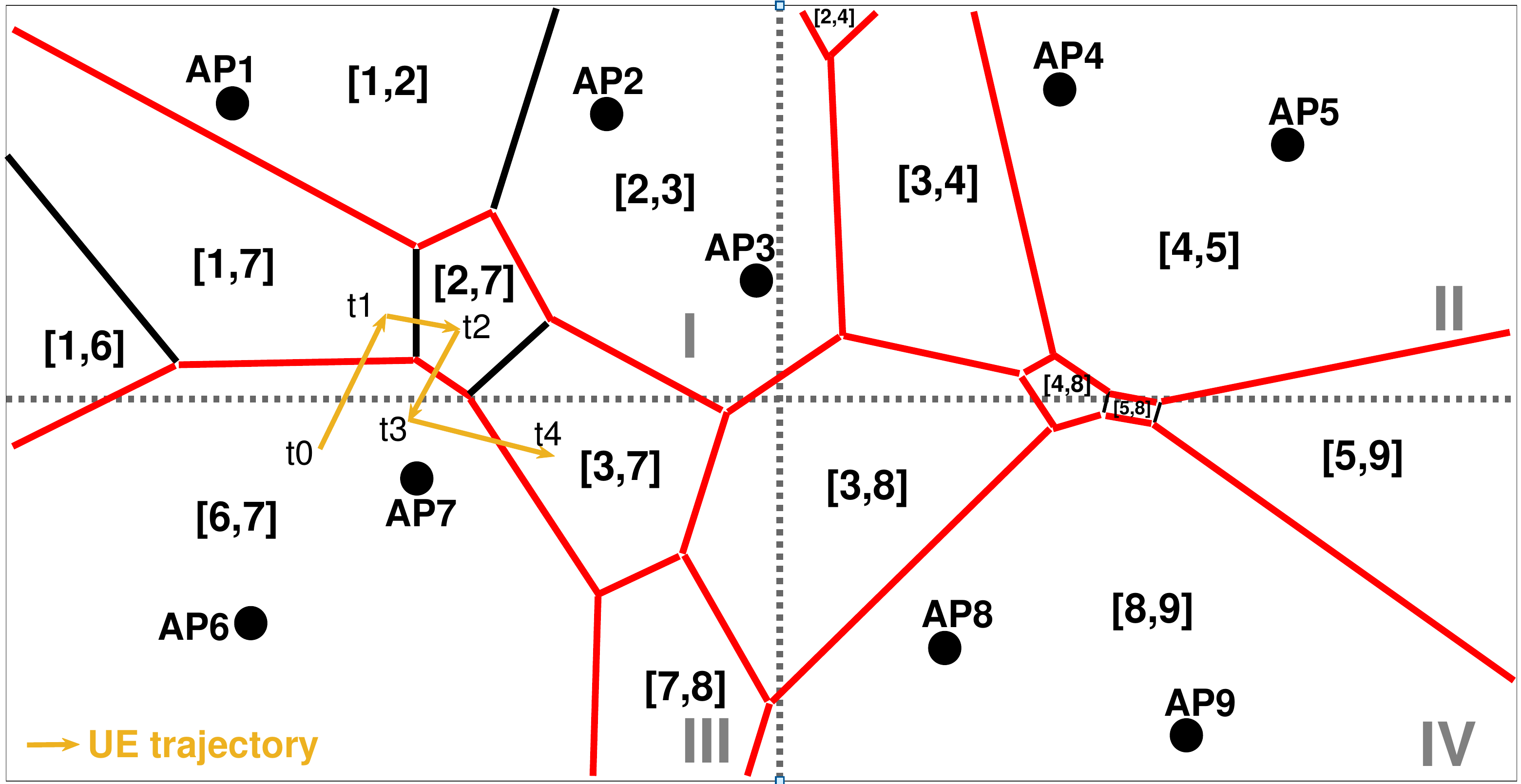} 
	\caption{Handover boundaries for the hybrid AP selection method. The gray dotted lines represent the border of CPU cluster areas. The solid (black and red) lines mark the edges of the $K^{th}$- order Voronoi cells. The red lines are the handover boundaries, corresponding to the modified $K^{th}$- order Voronoi cell boundary set in (\ref{T1}).}
	\label{boundary}
\end{figure}

An example of the handover boundary for the hybrid method when $K=2$ is illustrated in Fig \ref{boundary}. The gray dotted lines represent the borders of CPU cluster areas. For example, APs 1, 2, and 3 are in CPU cluster I, while APs 6 and 7 are in cluster III. The $2^{\text{nd}}$-order Poisson Voronoi cell region of APs $x$ and $y$ is labelled as $[x,y]$. When the UE is inside area $[x,y]$, the closest two APs for this UE are $x$ and $y$. The solid (black and red) lines mark the edges of the $K^{th}$- order Voronoi cells. A handover occurs when a new AP serving set is formed. In the hybrid method, this only happens when a newly selected $K^{th}$-closest AP belongs to a new (previously non-serving) CPU cluster. This is the case when the UE crosses the handover boundaries marked in red in Fig. \ref{boundary}, corresponding to the modified $K^{th}$- order Voronoi cell boundary set in (\ref{T1}). An example UE trajectory is also shown in Fig. \ref{boundary} as yellow arrows. From start time $t0$ to $t1$, the UE crosses a red line and the closet AP set changes from [6,7] to [1,7]. Since AP 1 belongs to CPU cluster I and APs 6 and 7 belong to III, the serving cluster set changes from cluster III to clusters I and III and a handover occurs. By contrast, from $t1$ to $t2$, the UE crosses a black line and only changes the closest AP set, since AP 1 and 2 are both in cluster I. Therefore, on the whole trajectory from $t0$ to $t4$, the UE crosses the red boundaries three times and there are in fact 3 handovers.

\subsection{Length Intensity}

Length intensity is the expected length of the handover boundary in a unit square \cite{optimizing}. The handover rate is derived through the length intensity of this boundary set, which is in turn derived through the area intensity of the extended boundary. The length intensity of $\mathbf{T}_{0K}^{(1)}$ is
\begin{equation}
\mu_{1}\left(\mathbf{T}_{0K}^{(1)}\right)=\mathbb{E}\left(\left|\mathbf{T}_{0K}^{(1)} \bigcap[0,1)^{2}\right|_{1}\right),
\label{uT1}
\end{equation}
where $|A|_1$ is the one-dimensional Lebesgue measure of $A$.

\noindent The $\Delta d$-extended boundary is defined as 
\begin{equation}
\mathbf{T}_{0K}^{(2)}(\Delta d)=\left\{\mathbf{y} \in \mathbb{R}^{2} \mid \exists \mathbf{x} \in \mathbf{T}_{0K}^{(1)}, \text { s.t. }|\mathbf{x}-\mathbf{y}|<\Delta d\right\}.
\label{T2}
\end{equation}

\noindent The area intensity of $\mathbf{T}_{0K}^{(2)} $ is the expected area in a unit square given by
\begin{equation}
\mu_{2}\left(\mathbf{T}_{0K}^{(2)}(\Delta d)\right)=\mathbb{E}\left(\left|\mathbf{T}_{0K}^{(2)}(\Delta d) \bigcap[0,1)^{2}\right|_{2}\right),
\label{uT2}
\end{equation}
where $|B|_2$ is the two-dimensional Lebesgue measure of $B$.

\noindent The area intensity is equal to the probability that an arbitrarily located point in the unit square is in this area. When the point is at the origin point $\mathbf{0}$, the area intensity for $\mathbf{T}_{0K}^{(2)} $ may be expressed as
\begin{equation}
\mu_{2}\left(\mathbf{T}_{0K}^{(2)}(\Delta d)\right)=\mathbb{P}\left(\mathbf{0} \in \mathbf{T}_{0K}^{(2)}(\Delta d)\right).
\label{uT2P}
\end{equation}

\noindent The probability in (\ref{uT2P}) may be rewritten as \cite{disjoint}
\begin{equation}
\begin{aligned}
\mathbb{P}&(\mathbf{0}\in \mathbf{T}_{0K}^{(2)}(\Delta d)) \\
&=\int_{0}^{\infty} \mathbb{P}(\mathbf{0} \in \mathbf{T}_{0K}^{(2)}(\Delta d) | R_{K}=r_{0})
 f_{K}(r_{0}) \mathrm{d} r_{0},
\end{aligned}
\label{P}
\end{equation}

\noindent where $R_K$ is the distance between $\mathbf{0}$ and its $K^{th}$ closest AP and $f_{K}(r_{0})$ is the probability density function of the distance between a UE and its $K^{th}$ closest AP, which is given by \cite{distance}
\begin{equation}
f_{K}(r)=\frac{2\left(\lambda \pi r^{2}\right)^{K}}{r \Gamma(K)} \exp \left(-\lambda \pi r^{2}\right).
\label{r}
\end{equation}


According to \cite{optimizing}, $\mathbf{0} \in \mathbf{T}_{0K}^{(2)}(\Delta d)$ only when $\mathbf{x}_{K}^{\prime}$ is located in the ring area
\begin{equation}
\begin{aligned}
S(\Delta d)&=\{(r, \theta) | r \geq r_{0} , |r^{2}-r_{0}^{2}| \\
&<2 \Delta d \sqrt{r_{0}^{2}+r^{2}-2 r_{0} r \cos \theta}\}.
\end{aligned}
\label{ring}
\end{equation}


\noindent The conditional probability in \eqref{P} may then be expressed as 
\begin{equation}
\begin{aligned}
\mathbb{P} &\left( \mathbf{0} \in \mathbf{T}_{0K}^{(2)}(\Delta d) \mid R_K =r_{0}\right) \\
&= \frac{1-e^{\lambda|S(\Delta d)|}}{|S(\Delta d)|} \cdot 
 \int_{s(\Delta d)} P\left(L, r_{0} \sqrt{2-2 \cos \theta}\right) r d r d \theta \\
&= 2 \Delta d r_{0} \lambda  
\int_{0}^{\pi} \sqrt{2-2 \cos (\theta)} P\left(L, r_{0} \sqrt{2-2 \cos (\theta)}\right) \mathrm{d} \theta \\
&+O(\Delta d^{2}),
\end{aligned}
\end{equation}

\noindent where $P$ denotes the probability that the two APs are not in the same CPU cluster. For length $L$ of the square CPU cluster and  distance $r$ between the two APs, $P(L, r)$ can be derived by the Buffon’s needle problem on a square grid \cite{disjoint}.

\subsection{Handover Rate}
From the above, we derive the area intensity of $\mathbf{T}_{0 K}^{(2)}$ as
\begin{equation}
\begin{aligned}
\mu_{2}\left( \mathbf{T}_{0 K}^{(2)}(\Delta d)\right) &=
\int_{0}^{\frac{\sqrt{2}}{2} L} \frac{4 \Delta d \lambda(\lambda \pi)^{K}}{\Gamma(k)} r_{0}^{2 K} e^{-\lambda \pi r_{0}^{2}} \cdot \\
& \int_{0}^{\pi} \sqrt{2-2 \cos \theta} P\left(L, r_{0} \sqrt{2-2 \cos \theta}\right) d \theta d r_{0} \\
&+ 2 \Delta d \int_{0}^{\pi} \sqrt{2-2 \cos \theta} d \theta \sqrt{\frac{\lambda}{\pi}}  \frac{\Gamma(K+1/2)}{\Gamma(K)} \\
&- \int_{0}^{\frac{\sqrt{2}}{2} L} \frac{4 \Delta d \lambda(\lambda \pi)^{K}}{\Gamma(K)} \cdot \\
& \int_{0}^{\pi} \sqrt{2-2 \cos \theta} d \theta \cdot r_{0}^{2 K} e^{-\lambda \pi r_{0}^{2}} d r_{0},
\end{aligned}
\label{H-exact}
\end{equation}

\noindent using which the exact numerical result for the handover rate is obtained.

Let us consider a CPU cluster of size $Q=\lambda L^2$. When $Q$ is large, i.e., there are many APs in one CPU cluster, $\lambda r_0^2$ is large and $e^{-\lambda \pi r_{0}^{2}} \rightarrow 0$. We can then substitute the expression for $P(L, r)$ in Eq. (43) in \cite{disjoint} for the case when $r < L$ into (\ref{H-exact}) and obtain the closed form expression
\begin{equation}
\mu_{2}\left(\mathbf{T}_{0 K}^{(2)}(\Delta d)\right) = \frac{8 K  2 \Delta d}{\pi L}-\frac{32 \cdot 2 \Delta d }{3 L^{2} \pi^{2} \sqrt{\lambda \pi}}  \frac{\left(K+\frac{1}{2}\right) \Gamma\left(K+\frac{1}{2}\right)}{\Gamma(K)}.
\label{H-appr}
\end{equation}

\noindent We note that for a large $K$, a larger $Q$ will be required for an accurate approximation; we will show in Sec. \ref{simH} that the accuracy is in fact decided by the ratio $Q/K$.

Then the length intensity of $\mathbf{T}_{0K}^{(1)}$ is derived as
\begin{equation}
\mu_{1}\left(\mathbf{T}_{0K}^{(1)}\right)=\lim _{\Delta d \rightarrow 0} \frac{\mu_{2}\left(\mathbf{T}_{0K}^{(2)}(\Delta d)\right)}{2 \Delta d}.
\end{equation}

For a mobile UE that moves with speed $v$ and direction $\theta$ in each mobility period, the CPU cluster handover rate is
\begin{equation}
{H}_{C} = E[v|sin\theta|] \times \mu_{1}\left(\mathbf{T}_{0K}^{(1)}\right)
=\frac{2}{\pi} \mu_{1}\left(\mathbf{T}_{0K}^{(1)}\right) v.
\end{equation}

Since APs are uniformly distributed in the area and the handover happens one disjoint CPU cluster at a time, the expectation of the number of AP handovers for one CPU cluster handover is $\lambda L^2$. Thus the AP handover rate is
\begin{equation}
{H}_{AP} = {H}_{C} \times \lambda L^2.
\label{C2AP}
\end{equation}

Therefore, the closed form approximations of the CPU cluster and AP handover rate, respectively, for the hybrid method are given by
\begin{equation}
H_{C}^{hyb}(K,L,\lambda,v)=\frac{16 K v}{\pi^{2} L}-\frac{64 v}{3 L^{2} \pi^{3} \sqrt{\lambda \pi}} \cdot \frac{\left(K+\frac{1}{2}\right) \Gamma\left(K+\frac{1}{2}\right)}{\Gamma(K)},
\label{Hc-app}
\end{equation}
\begin{equation}
H_{A P}^{hyb}(K,L,\lambda,v)=\frac{16 K \lambda L v}{\pi^{2}}-\frac{64 v \sqrt{\lambda}}{3 \pi^{3} \sqrt{\pi}} \cdot \frac{\left(K+\frac{1}{2}\right) \Gamma\left(K+\frac{1}{2}\right)}{\Gamma(K)}.
\label{Hap-app}
\end{equation}

\section{Mobility-Aware User Throughput Performance Model}
\label{SEmo}

\subsection{Handover Rates for Different AP Selection Methods}
The handover rates for the hybrid method are derived in \eqref{Hc-app} and \eqref{Hap-app}. For completeness, we list the handover rates for CoMP-JT and PUE as follows. For CoMP-JT, the CPU cluster handover rate is \cite{tutorial}
\begin{equation}
H_{C}^{CoMP}(v,L) = \frac{4}{\pi} \frac{v}{L},
\label{Hc-comp}
\end{equation}

\noindent and substituting (\ref{Hc-comp}) into (\ref{C2AP}), the AP handover rate is
\begin{equation}
H_{AP}^{CoMP}(v,\lambda, L) = \frac{4}{\pi} \lambda L v.
\label{Hap-comp}
\end{equation}

\noindent For PUE, the AP handover rate is \cite{optimizing}
\begin{equation}
H_{AP}^{PUE}(v,\lambda, K) = \frac{8 \Gamma(K + 1/2) \sqrt{\lambda}}{\Gamma(K) \pi \sqrt{\pi}} v.
\label{Hap-pue}
\end{equation}

%


\subsection{Mobility-Aware Spectral Efficiency}

Mobility impacts the throughput due to the delay of each handover \cite{mobilityaware}. When there is no handover, the per-UE downlink throughput is obtained by the model in Eq. (24) in \cite{cfvs} as $SE$ (bit/s/Hz). In a unit time, i.e., 1s, there will be $H$ handovers each corresponding to $d$ (s) delay, during which the communication link will not function. Therefore, for each time unit only $(1-Hd)$ of the $SE$ will be obtained.


Two types of handover delay are considered in this paper: control plane delay $d_1$ and intra-cluster delay $d_2$. For the PUE method, all APs are directly managed by the control plane; thus for each handover the delay will be $d_1$. For CoMP-JT and the hybrid method, the APs are managed in CPU clusters and the handover happens at the level of clusters. First there would be a coordination among CPUs incurring delay $d_1$, then the APs within the CPU cluster establish new connections each with delay $d_2$. The mobility-aware spectral efficiency $SE'$ is thus given by
\begin{equation}
SE^{\prime}=\left\{\begin{array}{l}
SE(1-H_{A P} d_{1}) \text {,\qquad \qquad for PUE } \\
SE(1-H_{C} d_{1}-H_{A P} d_{2}) \text {, for CoMP-JT, hybrid}
\end{array}\right.
\label{SEm}.
\end{equation}

\noindent If $(1-H_{AP}d_1)$ or $(1-H_C d_1 - H_{AP} d_2)$ are $\leq 0$, then this link is blocked and $SE'=0$. The control plane delay is usually in the order of $10^{-1}$ s, while the intra-cluster delay would be much smaller, in the order of $10^{-2}$ s \cite{compdelay, differdelay}.

\section{Results}
\label{sim}

\subsection{Hybrid Method Handover Rate Validation}
\label{simH}

\begin{figure}[!t] 
	\centering  
	\subfigure[CPU cluster handover rate]{
		\label{HvSpeed.sub.1}
		\includegraphics[width=0.48\linewidth]{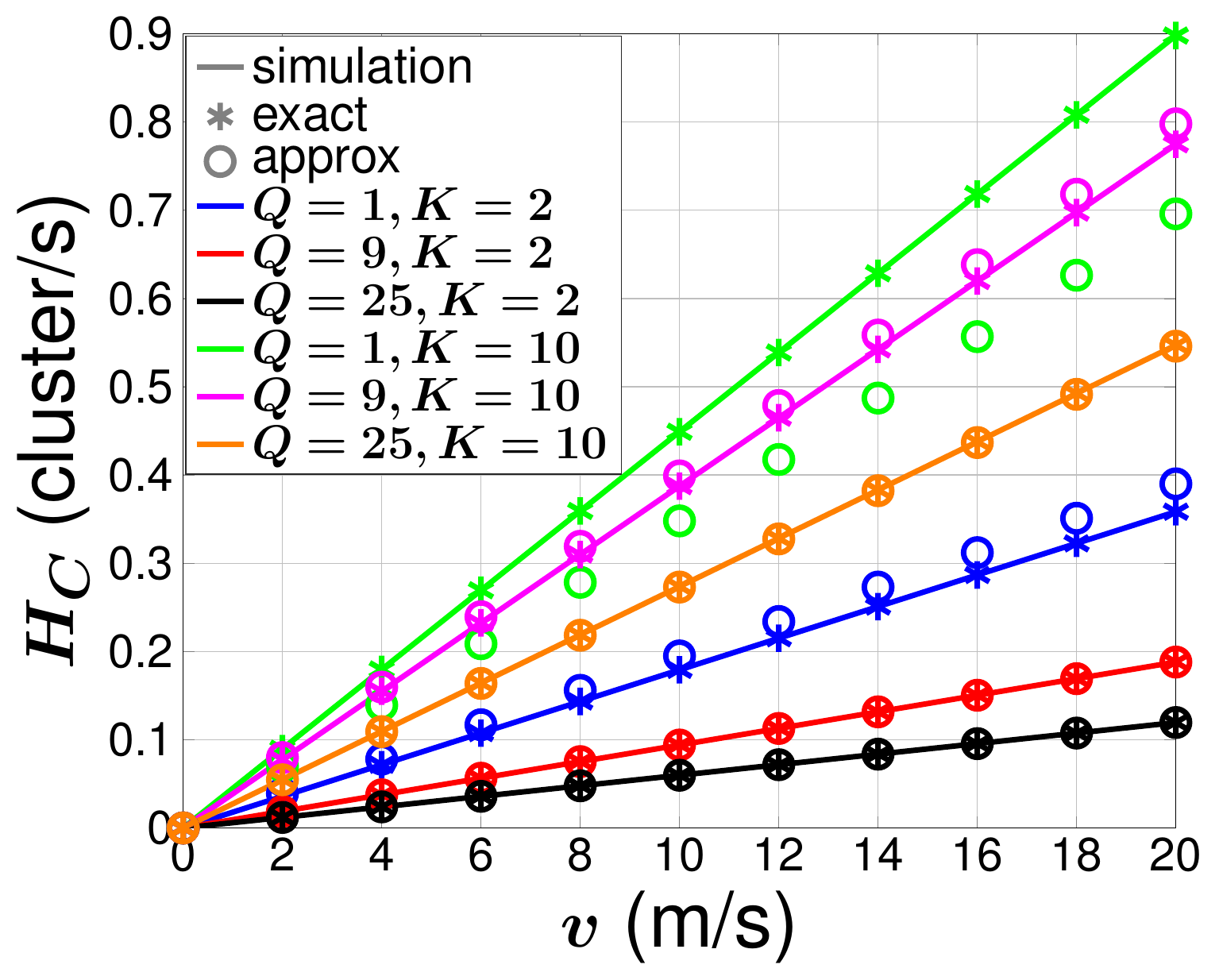}}
	\subfigure[AP handover rate]{
		\label{HvSpeed.sub.2}
		\includegraphics[width=0.48\linewidth]{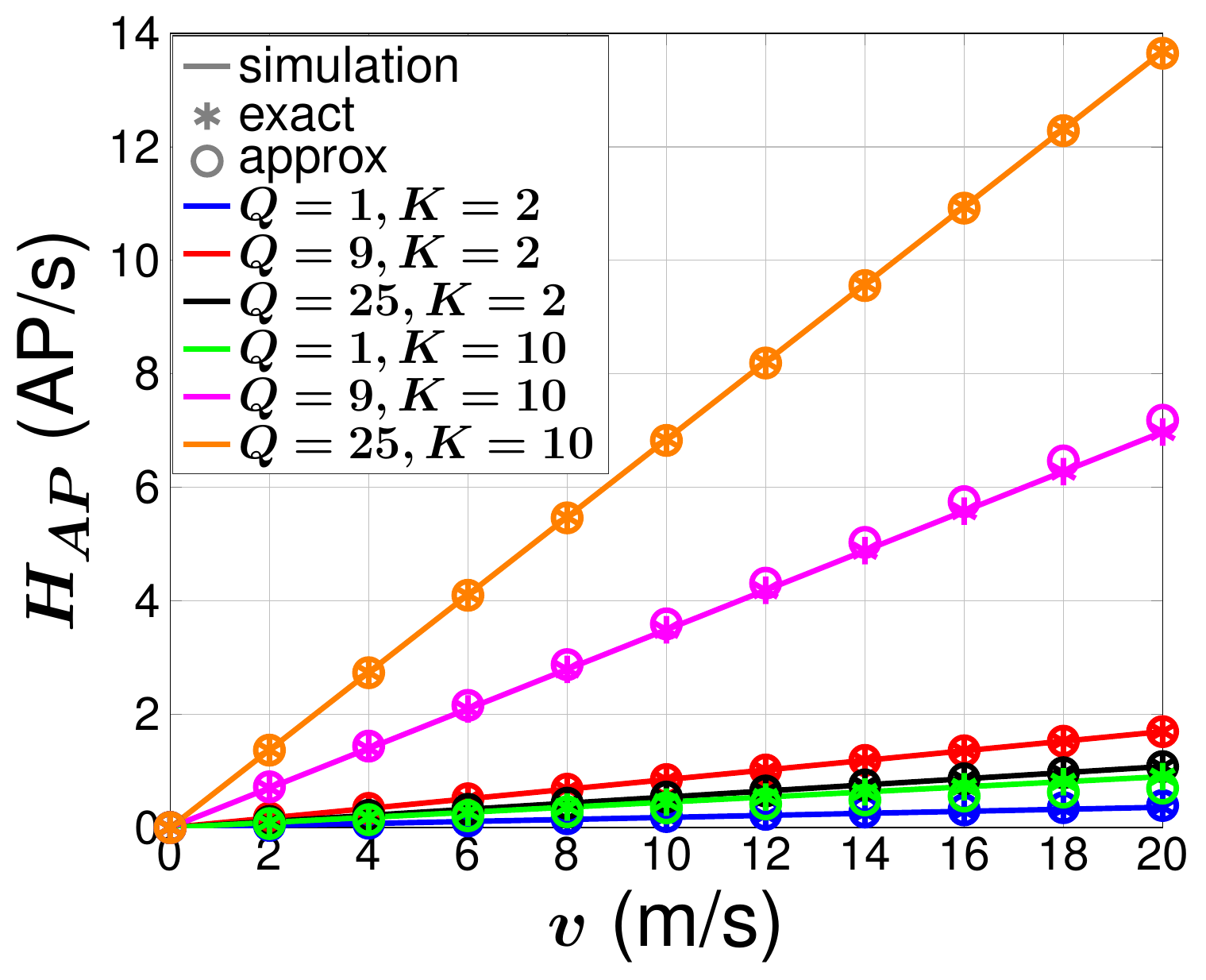}}
	\caption{Hybrid method handover rate vs. UE speed, for different $K$ and $Q$ ($\lambda=100 \; \text{APs/km}^2$). }
	\label{HvSpeed}
\end{figure}

In this section we validate the accuracy of our derived handover rate for the hybrid AP selection method used in scalable CF-mMIMO, against Monte Carlo simulations of 10,000 runs. For each run, APs and UEs are uniformly generated in a 20 km $\times$ 20 km area with density ratio $\lambda/\lambda_{U}=10$. The embedding technique in \cite{cfsim} is used to avoid border effects. Fig. \ref{HvSpeed} presents the CPU cluster and AP handover rates for the hybrid method versus UE speed for different values of $Q$ and $K$ with $\lambda=100 \; \text{APs/km}^2$, comparing our exact and closed form approximation analytical results against simulations. Figs. \ref{HvSpeed.sub.1} and \ref{HvSpeed.sub.2} confirm that our exact numerical results for the hybrid handover rate match the simulations for CPU cluster and AP handover rate, respectively. They also show that our closed form approximation results are accurate, unless $Q$ is small relative to $K$. This is because the serving AP set will then include many small clusters, thus effectively losing the feature of cluster-based AP selection underlying our analysis.

\begin{figure}[!t] 
	\centering  
	\subfigure[CPU cluster handover rate]{
		\label{QrK.sub.1}
		\includegraphics[width=0.48\linewidth]{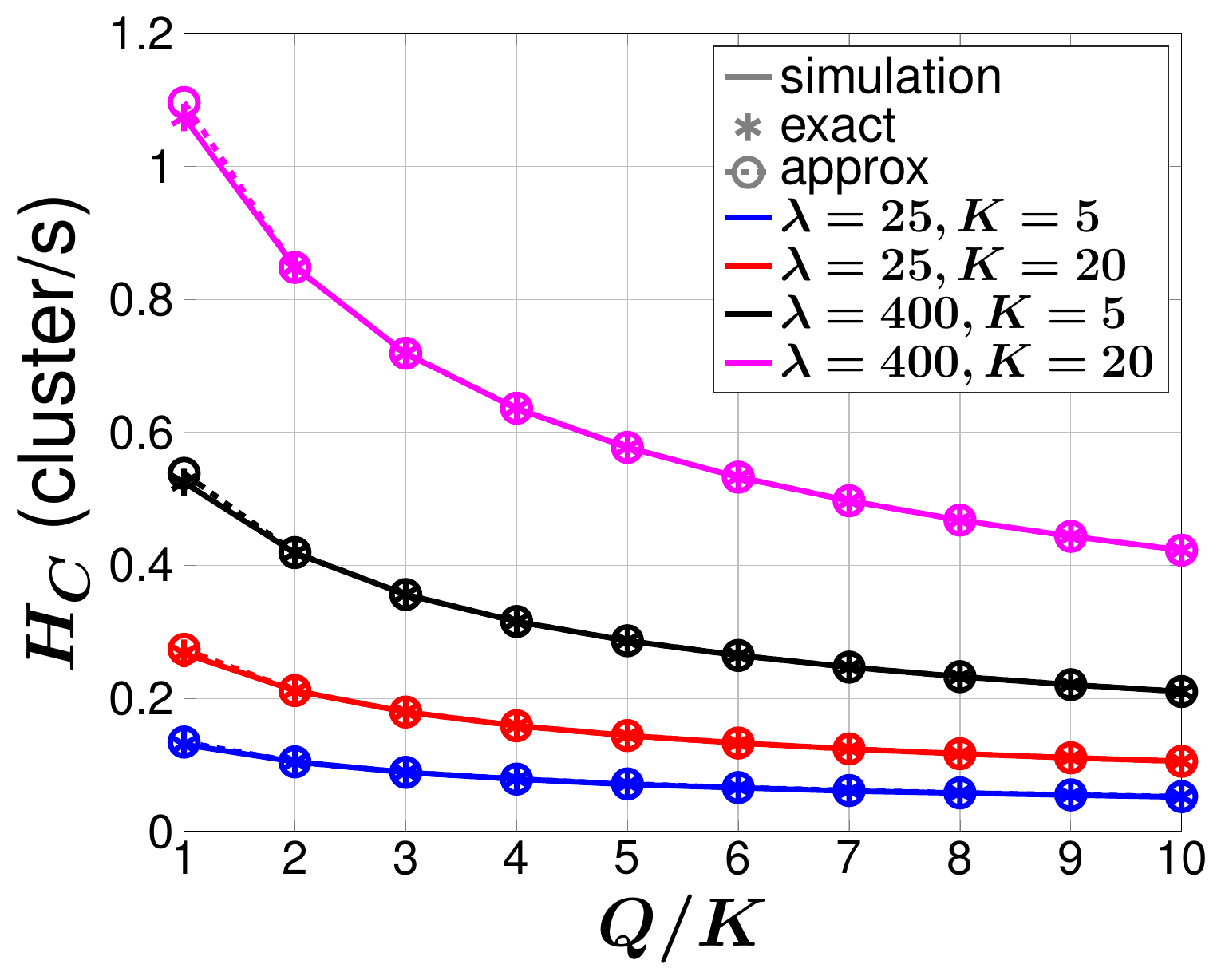}}
	\subfigure[AP handover rate]{
		\label{QrK.sub.2}
		\includegraphics[width=0.48\linewidth]{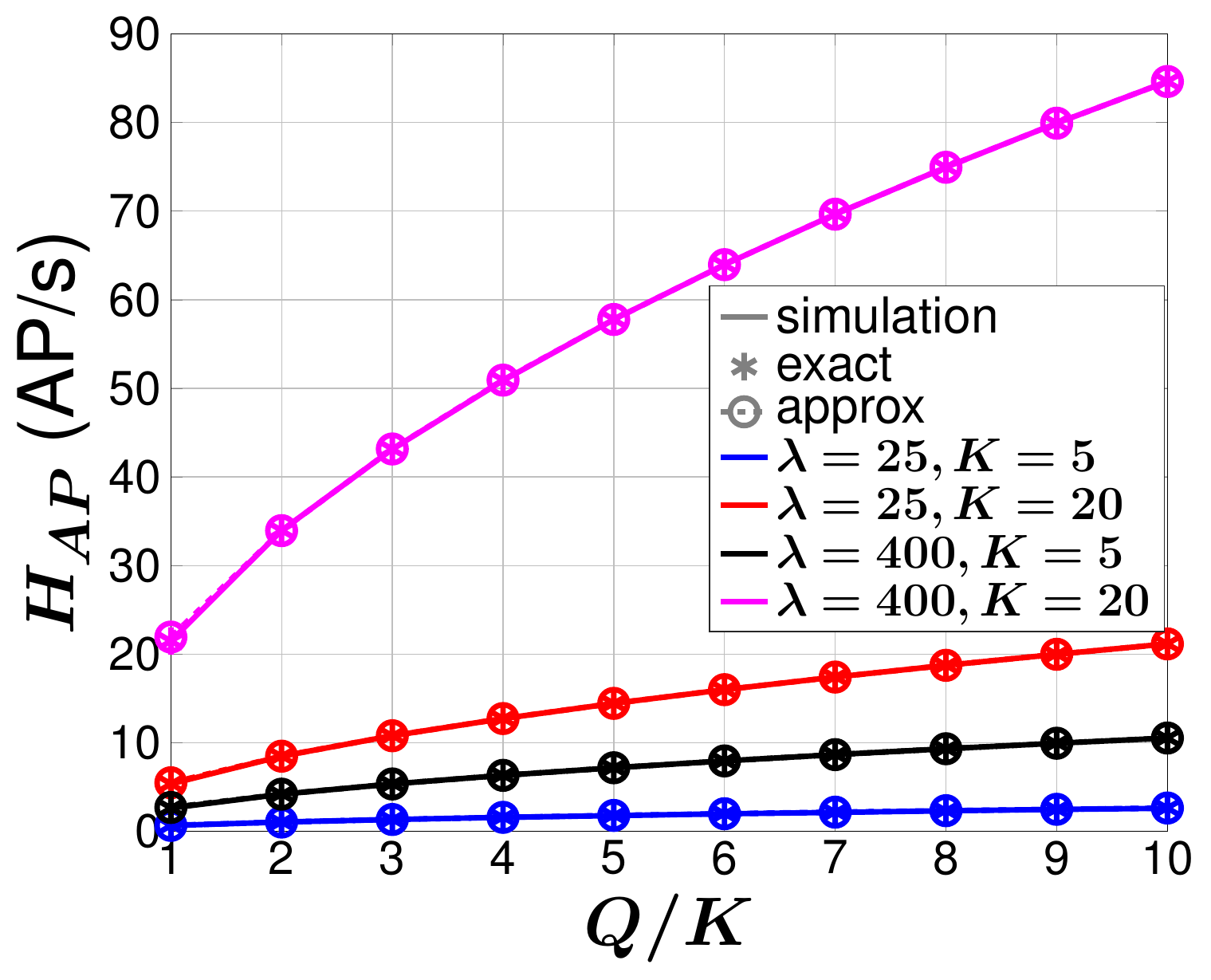}}
	\caption{Hybrid method handover rate vs. $Q/K$ ratio, for different $\lambda \;  (\text{AP/km}^2)$ and $K$ ($v=10$ m/s).}
	\label{QrK}
\end{figure}

In Fig. \ref{QrK} we observe the handover rate versus $Q/K$ ratio, at a fixed UE speed of 10 m/s. Fig. \ref{QrK.sub.1} shows that the CPU cluster handover rate decreases when $Q/K$ increases, whereas Fig. \ref{QrK.sub.2} shows the opposite trend for the AP handover rate. This is because cluster size grows when $Q/K$ increases, which leads to fewer cluster handovers but more APs changing connections in one handover. In both cases, for $Q/K \ge 2$ the approximation results match well with simulations. Importantly, the accuracy of our closed form approximation is invariant with respect to $\lambda$ and $K$. We emphasize that in practice, scalable CF-mMIMO would logically contain a relatively large number of APs per CPU cluster. Therefore, $Q/K \ge 2$ is a reasonable assumption for practical deployments, where our closed form expression provides accurate results and will be used in our mobility-aware performance analysis in the sequel.

\subsection{Mobility-Aware SE Performance}
\label{simSE}
In this section we compare the mobility-aware SE performance of the hybrid, CoMP-JT, and PUE AP selection methods for distributed MIMO. The static SE results are generated by the model in \cite{cfvs} with the same Monte Carlo settings as in Sec. \ref{simH}, and the mobility-aware SE model in \eqref{SEm} is used to quantify the mobility-aware performance. For a fair comparison, we observe the SE performance for a fixed average serving AP number ($N$) for each AP selection method, so that the serving AP set will have the same size but different membership: for CoMP-JT, the UE is always served by one CPU cluster and $Q=N$; for PUE, the serving set is formed by the closest $K$ APs, thus $K=N$; and for the hybrid method, the serving set is decided by both $Q$ and $K$. We choose $N=35$ and for hybrid the corresponding combinations $\{K,Q \}=\left\lbrace \{3,25\},\{5,20\},\{7,17\} \right\rbrace$. We consider the control plane delay of $d_1 = \{0.7 \; \text{s}, 0.1 \; \text{s}\}$ \cite{compdelay} and the intra-cluster delay of $d_2 = 0.02 \; \text{s}$ \cite{differdelay}.

\begin{figure*}[!t] 
	\centering  
	\subfigure[$\lambda = 400 \; \text{AP/km}^2$, $d_1$=0.7 s]{
		\label{medianD.sub.1}
		\includegraphics[width=0.23\linewidth]{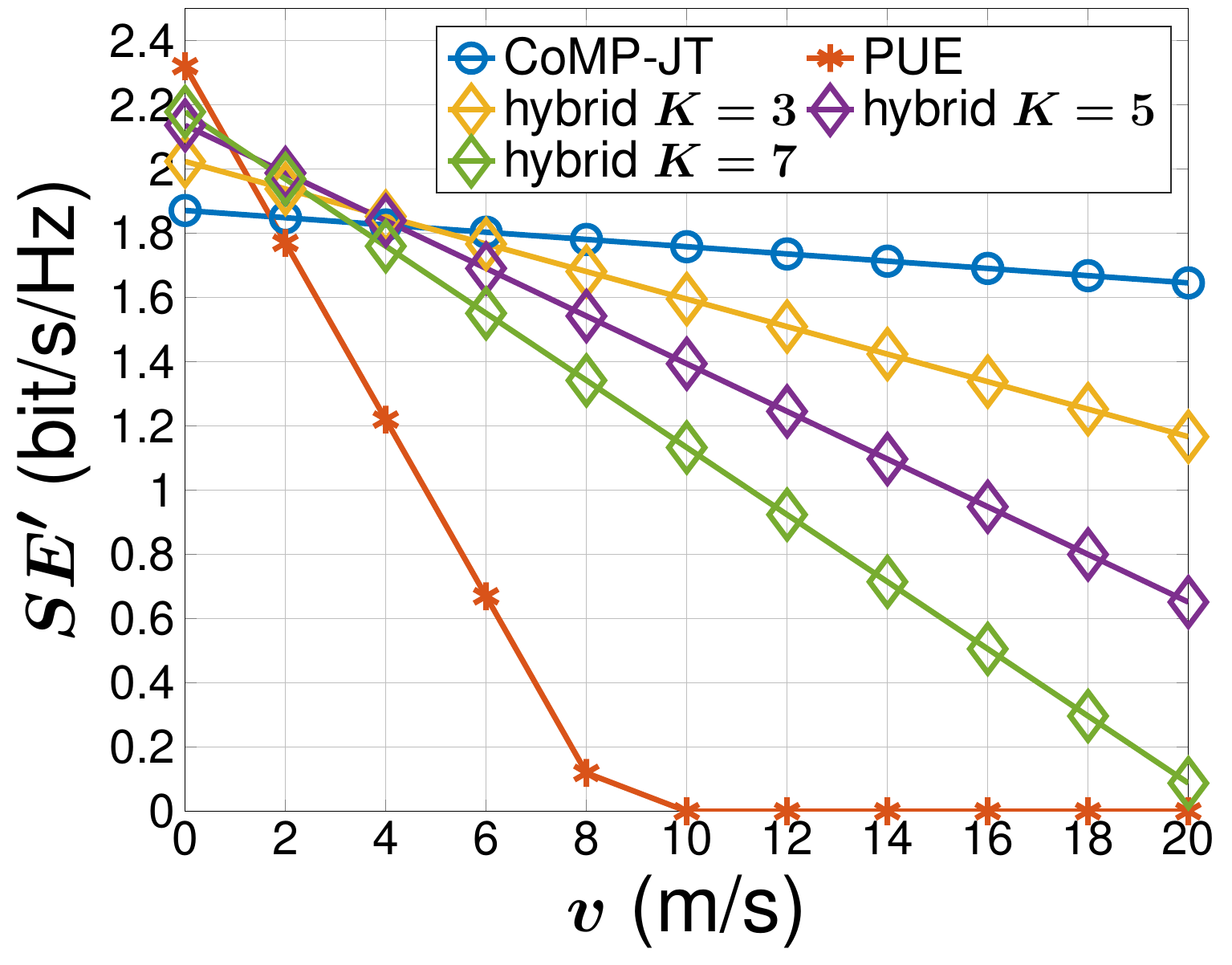}}	
	\subfigure[$\lambda = 25 \; \text{AP/km}^2$, $d_1$=0.7 s]{
		\label{medianD.sub.2}
		\includegraphics[width=0.23\linewidth]{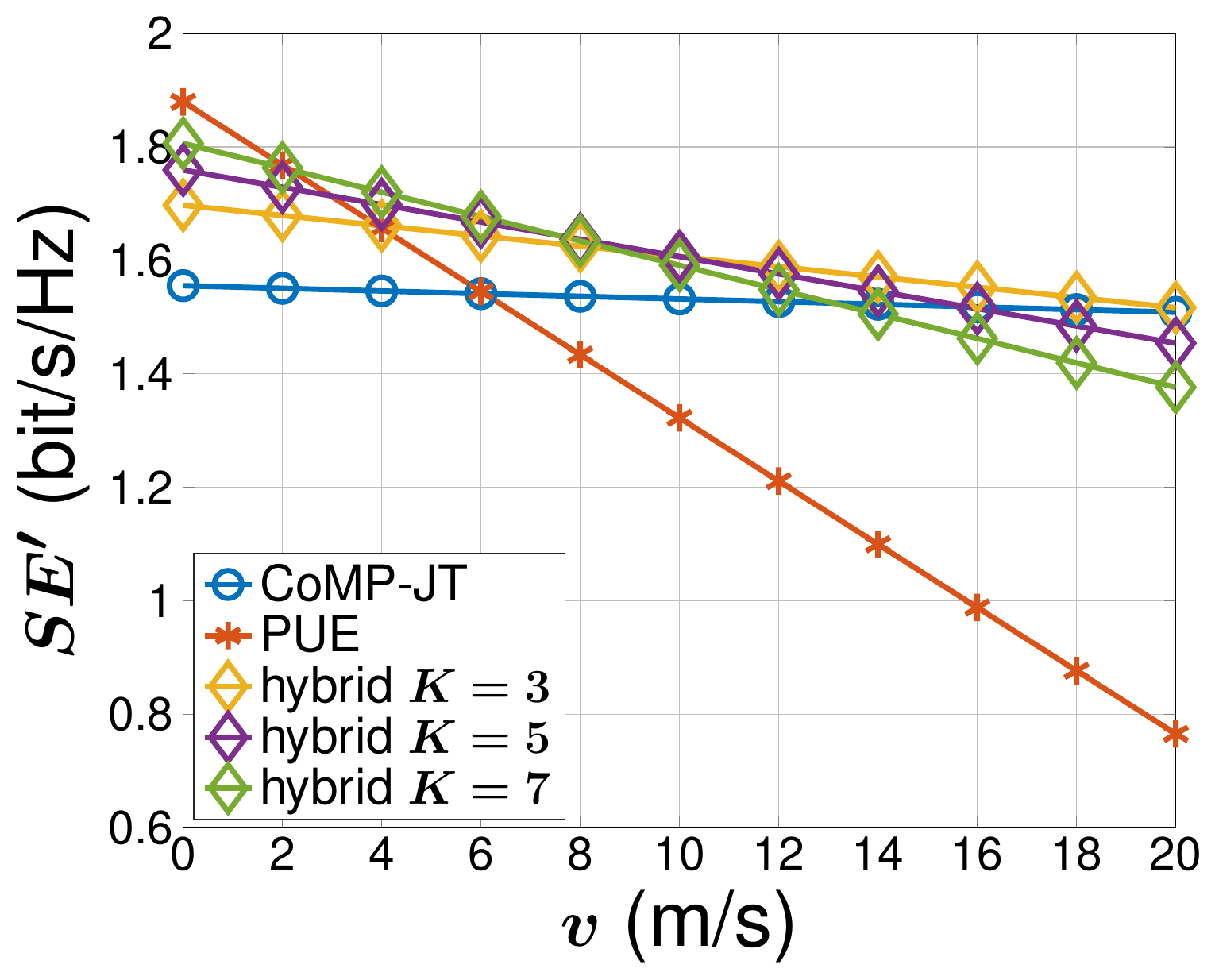}}
	\subfigure[$\lambda = 400 \; \text{AP/km}^2$, $d_1$=0.1 s]{
		\label{medianD.sub.3}
		\includegraphics[width=0.23\linewidth]{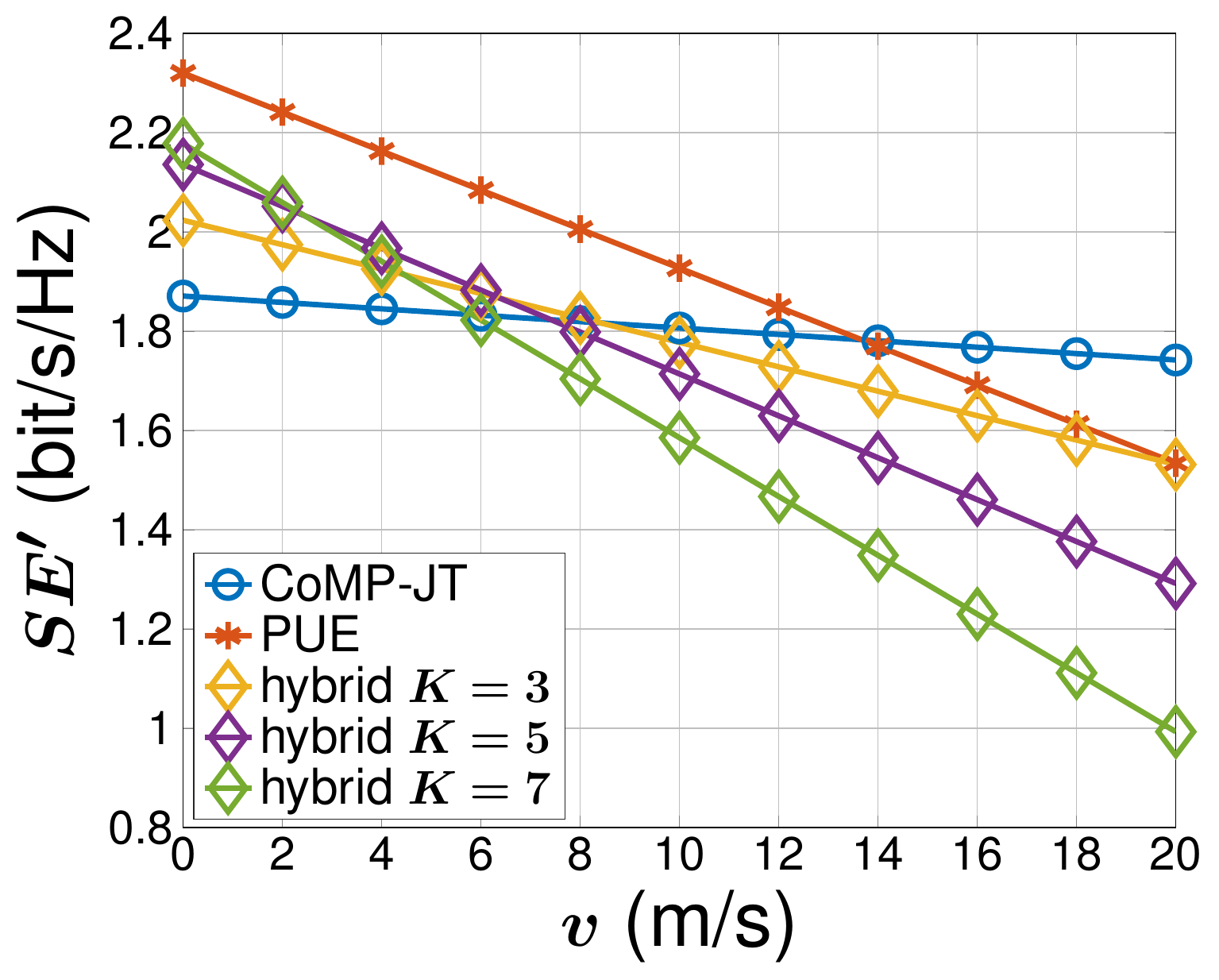}}
	\subfigure[$\lambda = 25 \; \text{AP/km}^2$, $d_1$=0.1 s]{
		\label{medianD.sub.4}
		\includegraphics[width=0.23\linewidth]{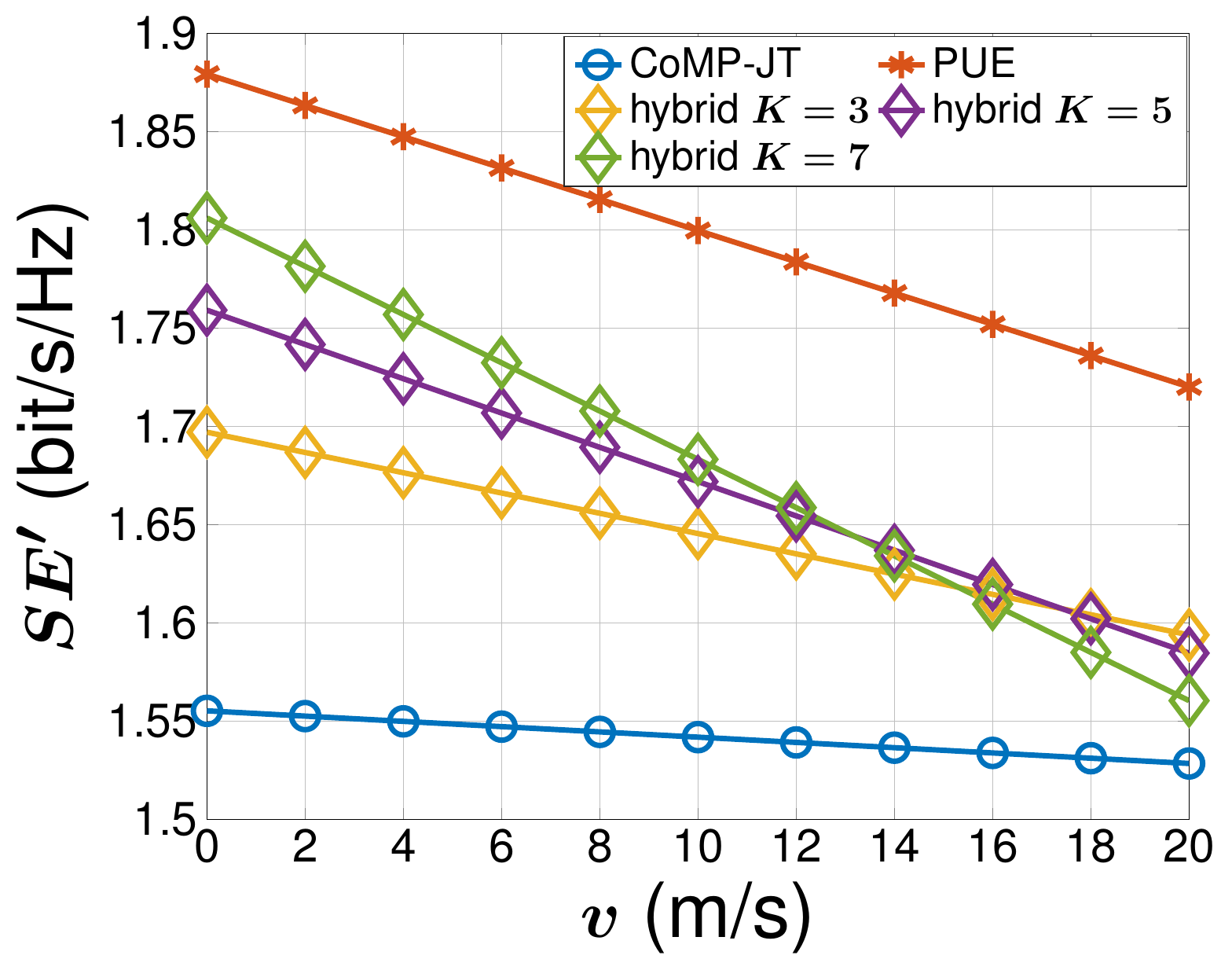}}
	\caption{Median mobility-aware SE performance vs. UE speed, for different AP selection methods, AP densities, and control plane delays.}
	\label{medianD}
\end{figure*}

Fig. \ref{medianD} presents the median SE performance versus UE speed for the three AP selection methods. With the increase of UE mobility, the mobility-aware SE decreases due to the delay of handovers. Fig. \ref{medianD.sub.1} shows that in a dense network with a high control plane delay, the hybrid method slightly outperforms CoMP-JT and PUE in median SE when the UE is at walking speed ($v \approx 3 \; \text{m/s}$), while Fig. \ref{medianD.sub.2} shows that this advantage extends up to $v \approx 11 \; \text{m/s}$ when AP density is low. Moreover, the hybrid method with smaller $K$ performs better at high mobility due to the correspondingly larger $Q$ and thus fewer cluster handovers. By contrast, for a low control plane delay, the hybrid method loses its significant advantage over PUE at low mobility and against CoMP-JT at high mobility, as evident from a comparison of Figs. \ref{medianD.sub.1} and \ref{medianD.sub.2} against Figs. \ref{medianD.sub.3} and \ref{medianD.sub.4}, respectively. This is because for PUE, the low handover delay does not harm the high SE provided by APs with the best channel condition even with a large number of handovers, whereas CoMP-JT has a relatively stable performance even at high mobility due to its low handover rate.

\begin{figure*}[!t] 
	\centering  
	\subfigure[$\lambda = 400 \; \text{AP/km}^2$, $d_1$=0.7 s]{
		\label{95D.sub.1}
		\includegraphics[width=0.23\linewidth]{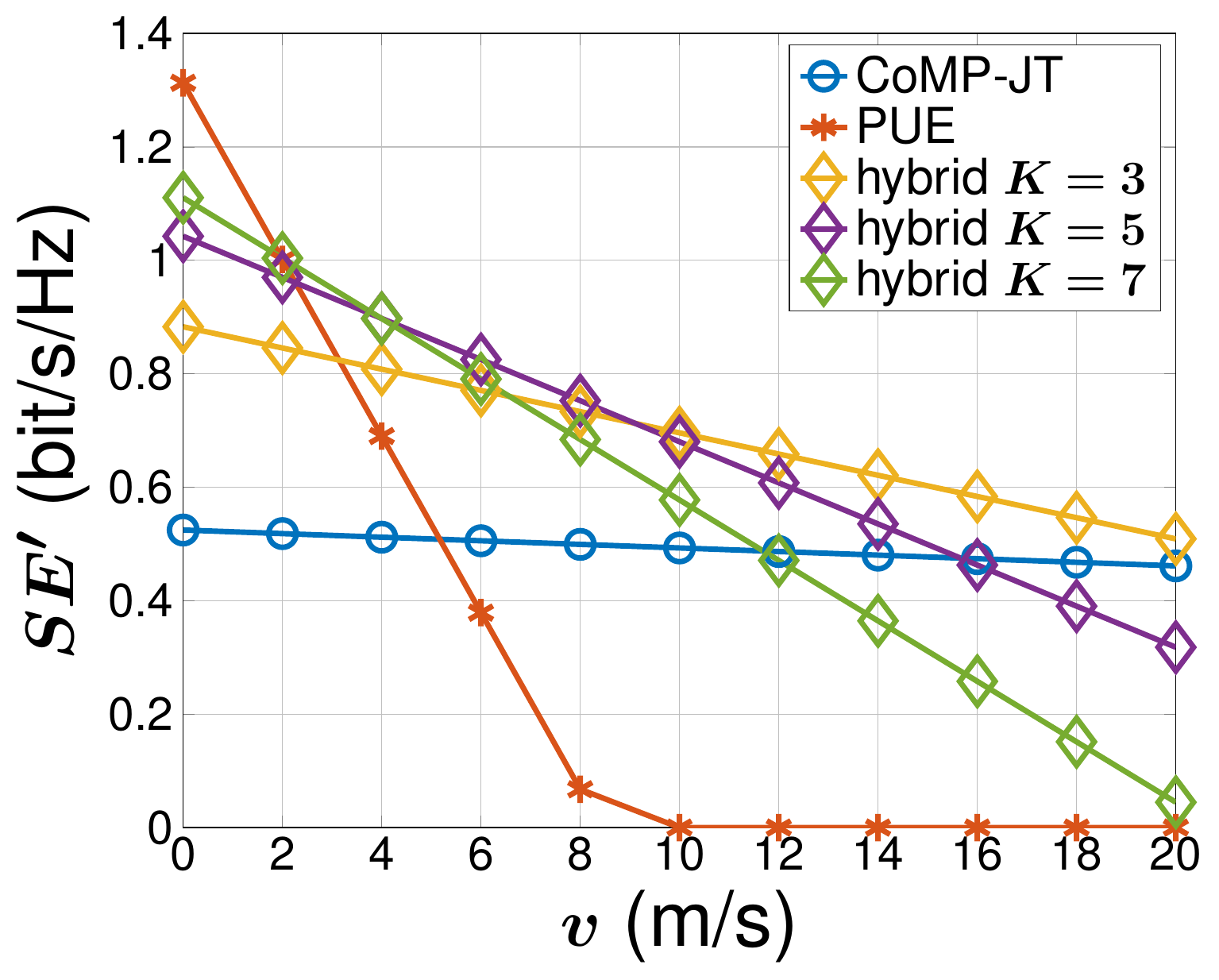}}
	\subfigure[$\lambda = 25 \; \text{AP/km}^2$, $d_1$=0.7 s]{
		\label{95D.sub.2}
		\includegraphics[width=0.23\linewidth]{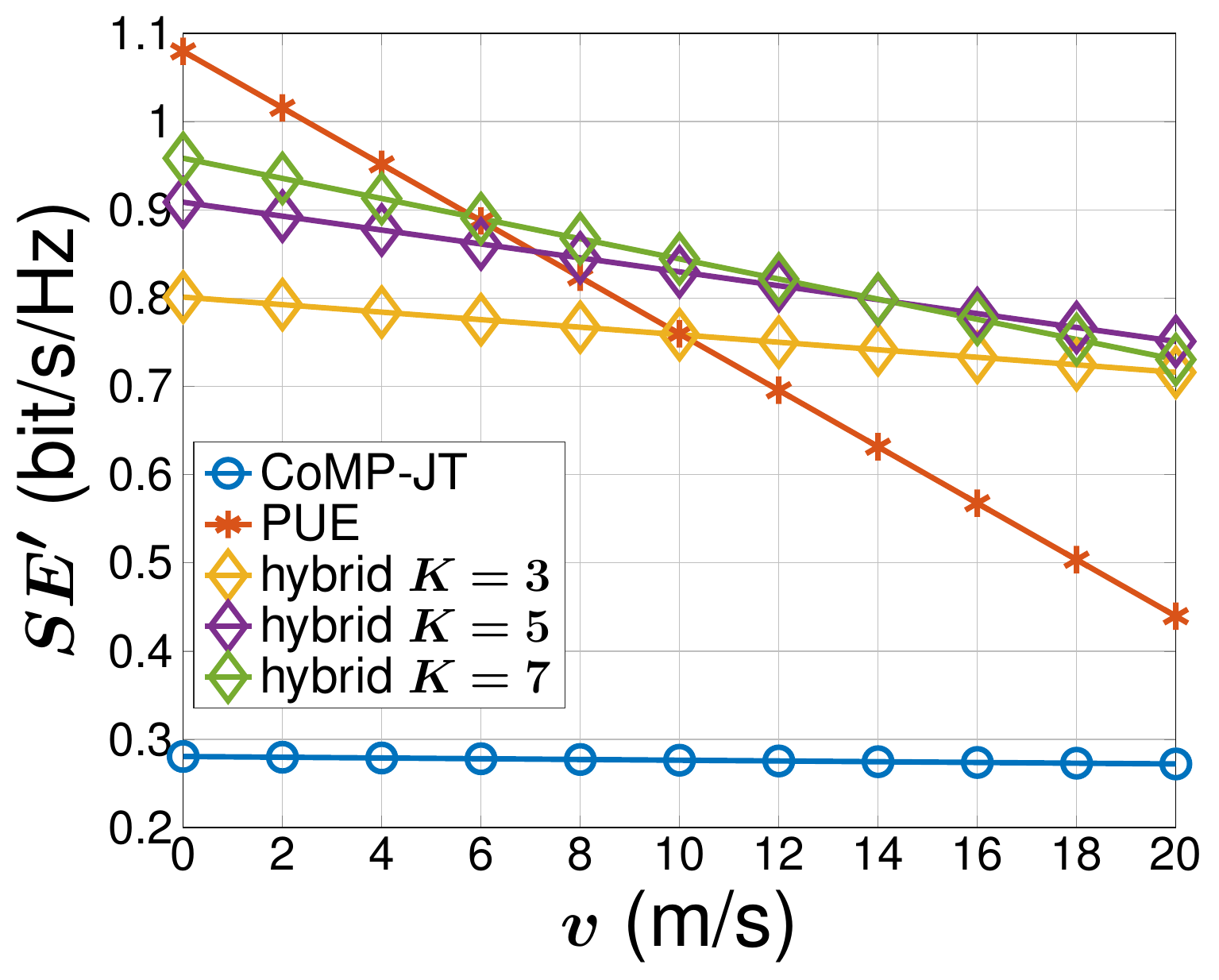}}
	\subfigure[$\lambda = 400 \; \text{AP/km}^2$, $d_1$=0.1 s]{
		\label{95D.sub.3}
		\includegraphics[width=0.23\linewidth]{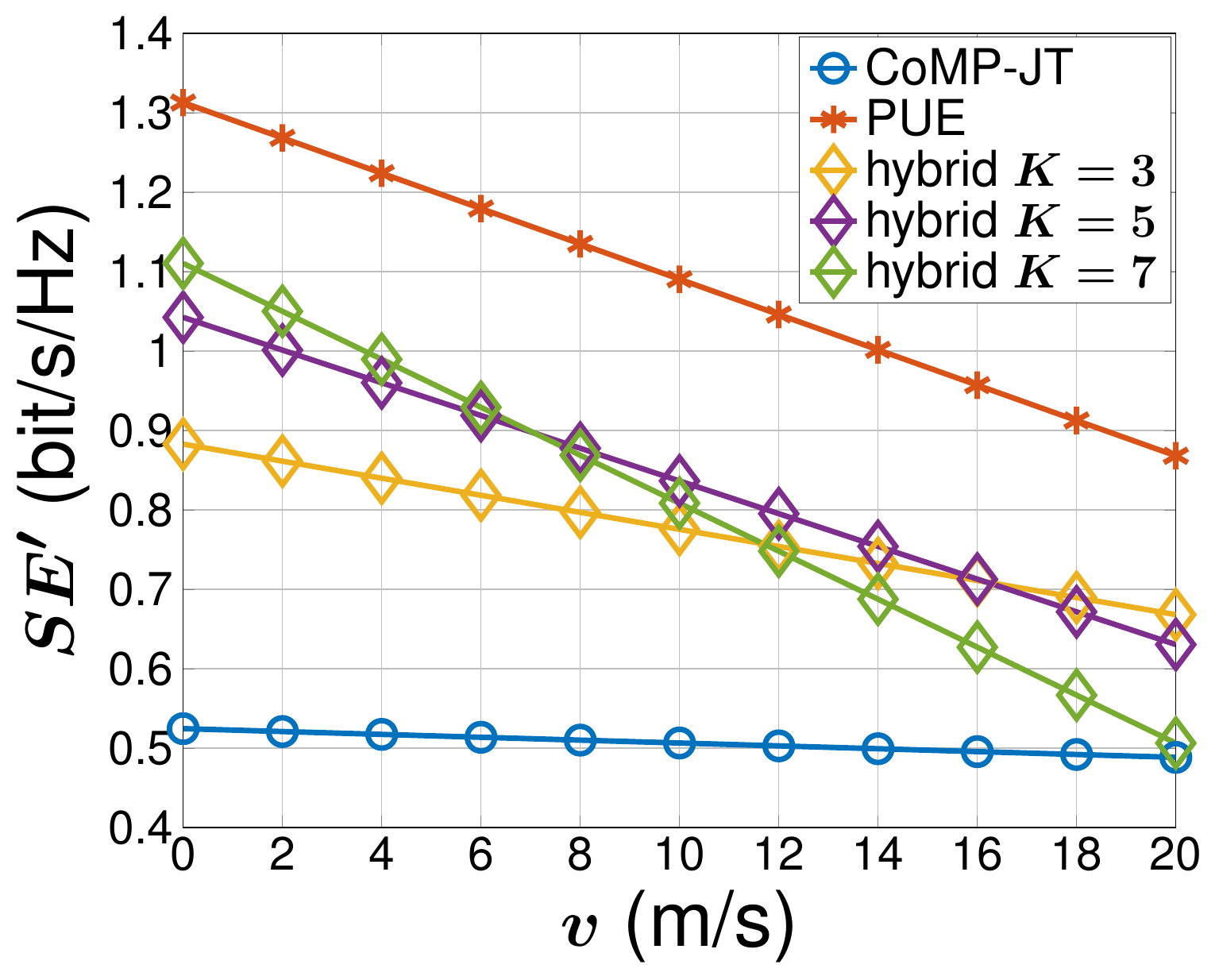}}
	\subfigure[$\lambda = 25 \; \text{AP/km}^2$, $d_1$=0.1 s]{
		\label{95D.sub.4}
		\includegraphics[width=0.23\linewidth]{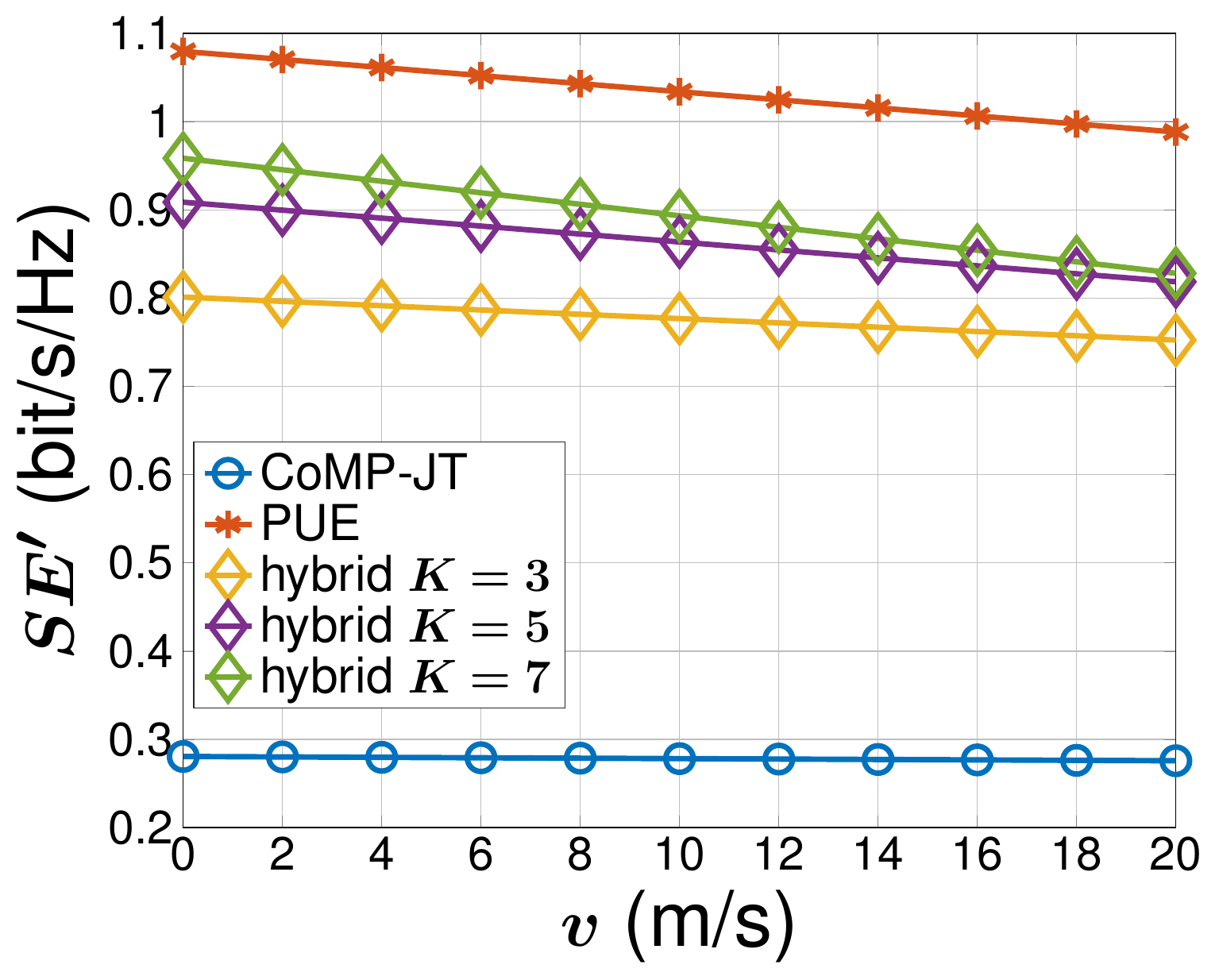}}

	\caption{95\%-likely mobility-aware SE performance vs. UE speed, for different AP selection methods, AP densities, and control plane delays.}
	\label{95D}
\end{figure*}

\begin{figure}[!t] 
	\centering  
	\subfigure[$\lambda = 400 \; \text{AP/km}^2$]{
		\label{Hrate.sub.1}
		\includegraphics[width=0.48\linewidth]{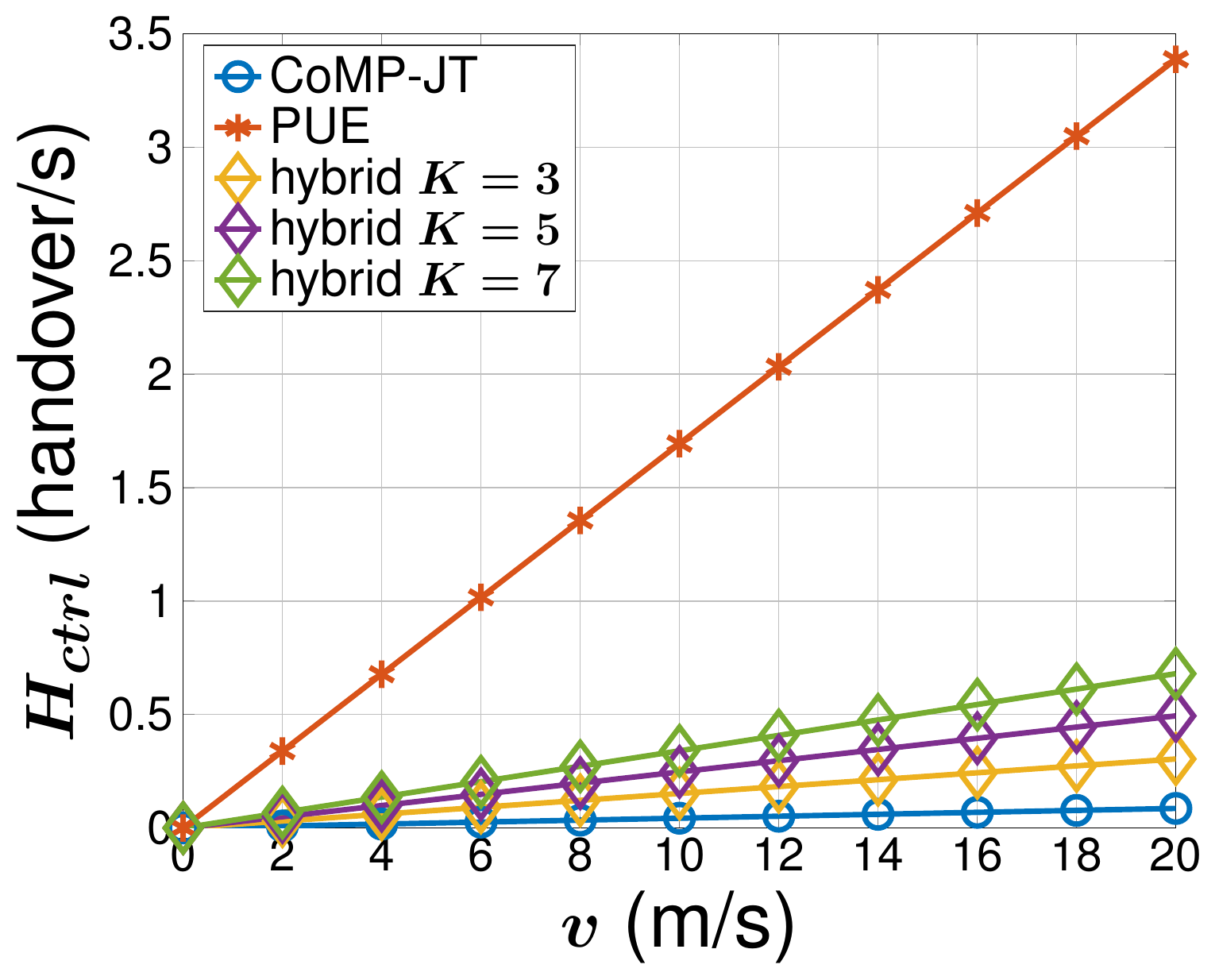}}
	\subfigure[$\lambda = 25 \; \text{AP/km}^2$]{
		\label{Hrate.sub.2}
		\includegraphics[width=0.48\linewidth]{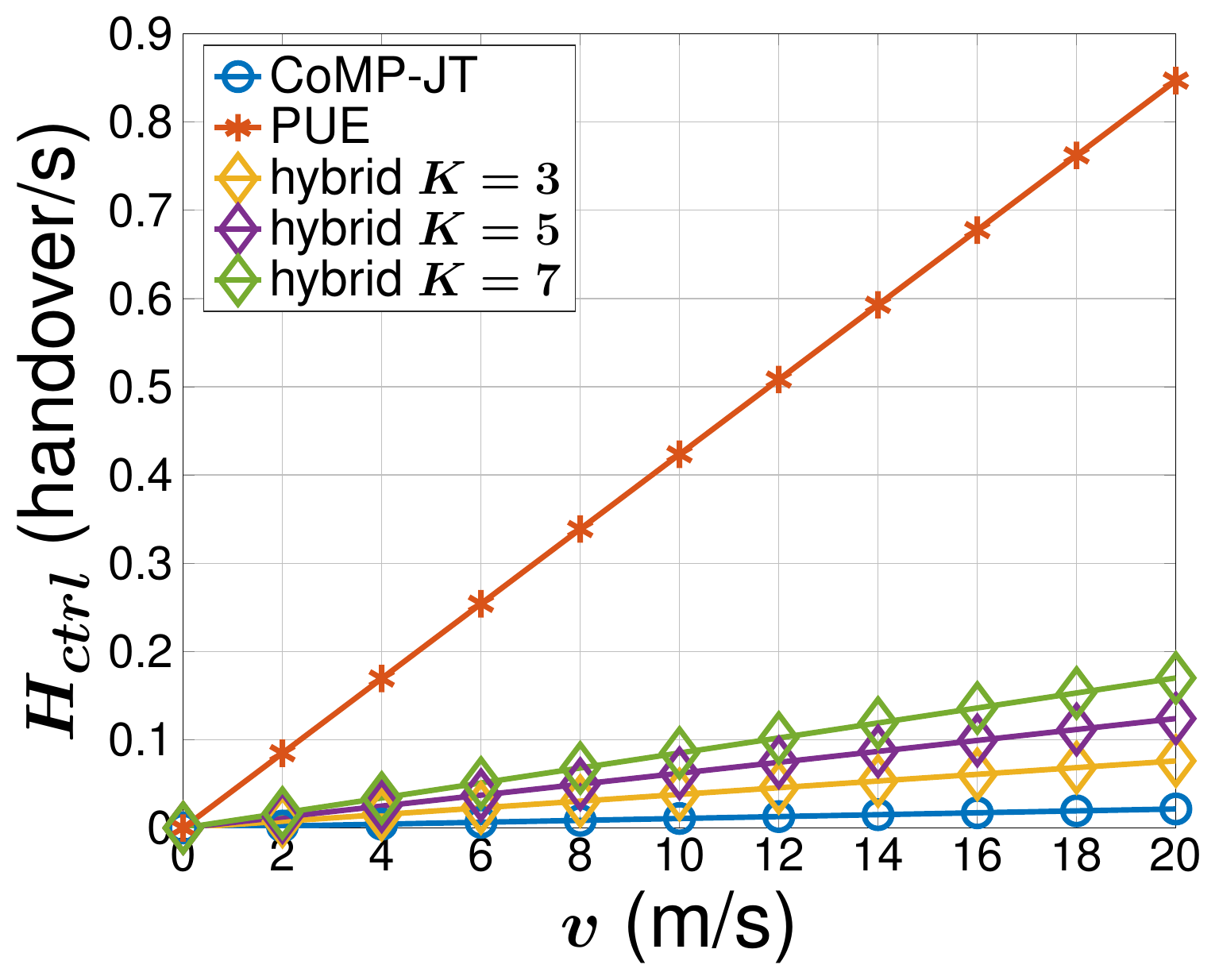}}
	\caption{Control plane handover rate vs. UE speed, for different AP selection method and AP densities.}
	\label{Hrate}
\end{figure}

Fig. \ref{95D} presents the 95\%-likely SE performance versus UE speed under the same conditions as in Fig. \ref{medianD}. For these \mbox{worst-served} UEs in all scenarios, PUE and the hybrid method significantly outperform CoMP-JT in the static network due to cluster-edge effect mitigation. Figs. \ref{95D.sub.1} and \ref{95D.sub.2} show that the hybrid method outperforms PUE and \mbox{CoMP-JT} for the worst-served UEs over a wider range of UE speeds than for the median, whereas Figs. \ref{95D.sub.3} and \ref{95D.sub.4} show that with small $d_1$, PUE maintains the highest SE in both high and low AP density. However, we emphasize that this superior SE performance of PUE comes at the cost of significantly higher control-plane signalling. This is illustrated by Fig.\ref{Hrate}, which shows that the control plane handover rate $H_{ctrl}$ for PUE is significantly higher than CoMP-JT and hybrid in both high and low density, especially at high mobility. This is because for PUE, all APs directly connect to the control plane, i.e., $H_{ctrl}=H_{AP}$, where for CoMP-JT and hybrid only CPU cluster handover is processed at the control plane, i.e., $H_{ctrl}=H_C$. This confirms that PUE is less attractive in practice than scalable CF-mMIMO, especially at high mobility.

Importantly, Fig. \ref{95D.sub.2} also shows that when AP density is low, the hybrid method (for all $K$) achieves a significantly higher SE than PUE and CoMP-JT from $v \approx 10 \; \text{m/s}$ (cycling speed) to $v \approx 20 \; \text{m/s}$ (urban driving speed). However, comparing Figs. \ref{medianD.sub.2} and Fig. \ref{95D.sub.2}, we observe that at high mobility the improvement for the worst-served UEs given by the hybrid method comes at the cost of reduced median SE. This trend is also observed in Figs. \ref{medianD.sub.3} and Fig. \ref{95D.sub.3}. Namely, with high AP density and low control delay -- which is the trend for future networks -- the hybrid method for scalable CF-mMIMO shows a trade-off between superior SE for worst-served UEs versus \mbox{CoMP-JT} and a substantially degraded median SE, even at moderate mobility ($v \textgreater 8 \; \text{m/s}$).

\section{Conclusions}
\label{conclude}
We derived exact expressions and closed form approximations for the CPU cluster and AP handover rate for scalable \mbox{CF-mMIMO} with hybrid AP selection. Simulation results confirmed that when $Q/K \ge 2$, the closed form expression is accurate. Based on our handover rate analysis, we compared the mobility-aware SE performance of the hybrid AP selection method for scalable CF-mMIMO against the network-centric CoMP-JT and the user-centric PUE. Our results showed that for future ultra-dense networks with low control delay, the hybrid method exhibits an important \mbox{trade-off} between the throughput of worst-served UEs and the median, for moderate to high mobility. Our ongoing work is studying in detail the impact of signalling overhead for a more comprehensive analysis of the mobility performance of distributed MIMO.
%
%
%
%
%
%
%
%
%
\ifCLASSOPTIONcaptionsoff
  \newpage
\fi
%
%
%
%
%
\bibliographystyle{IEEEtran}
\bibliography{IEEEabrv,IEEEexample}
\end{document}